\RequirePackage[switch]{lineno_babar_revtex}
\documentclass[twocolumn,showpacs,aps,prl,superscriptaddress,letterpaper,nofootinbib]{revtex4}
\catcode`\@=11\relax
\def\@makecol{%
 \setbox\@outputbox\vbox{%
  \boxmaxdepth\@maxdepth
 \protected@write\@auxout{}{%
 \string\@LN@col{\@ifnum{\pagegrid@cur=\@ne}{1}{2}}
      }%
  \@tempdima\dp\@cclv
  \unvbox\@cclv
  \vskip-\@tempdima
 }%
 \xdef\@freelist{\@freelist\@midlist}\global\let\@midlist\@empty
 \@combinefloats
 \@combineinserts\@outputbox\footins
  \set@adj@colht\dimen@
  \count@\vbadness
  \vbadness\@M
  \setbox\@outputbox\vbox to\dimen@{%
   \@texttop
   \dimen@\dp\@outputbox
   \unvbox\@outputbox
   \vskip-\dimen@
   \@textbottom
  }%
  \vbadness\count@
 \global\maxdepth\@maxdepth
}%
\def\balance@two#1#2{%
\outputdebug@sw{{\tracingall\scrollmode\showbox#1\showbox#2}}{}%
 \setbox\@ne\vbox{%
  \@ifvoid#1{}{%
   \unvcopy#1\recover@footins
   \@ifvoid#2{}{\marry@baselines}%
  }%
  \@ifvoid#2{}{%
   \unvcopy#2\recover@footins
  }%
 }%
 \dimen@\ht\@ne\divide\dimen@\tw@
 \dimen@i\dimen@
 \vbadness\@M
 \vfuzz\maxdimen
 \loopwhile{%
  \dimen@i=.5\dimen@i
  \outputdebug@sw{\saythe\dimen@\saythe\dimen@i\saythe\dimen@ii}{}%
  \setbox\z@\copy\@ne\setbox\tw@\vsplit\z@ to\dimen@
  \setbox\z@ \vbox{%
 \protected@write\@auxout{}{%
 \string\@LN@col{\@ifnum{\pagegrid@cur=\@ne}{1}{2}}
      }%
   \unvcopy\z@
   \setbox\z@\vbox{\unvbox\z@ \setbox\z@\lastbox\aftergroup\vskip\aftergroup-\expandafter}\the\dp\z@\relax
  }%
  \setbox\tw@\vbox{%
   \unvcopy\tw@
   \setbox\z@\vbox{\unvbox\tw@\setbox\z@\lastbox\aftergroup\vskip\aftergroup-\expandafter}\the\dp\z@\relax
  }%
  \dimen@ii\ht\tw@\advance\dimen@ii-\ht\z@
  \@ifdim{\dimen@i>.5\p@}{%
   \advance\dimen@\@ifdim{\dimen@ii<\z@}{}{-}\dimen@i
   \true@sw
  }{%
   \@ifdim{\dimen@ii<\z@}{%
    \advance\dimen@\tw@\dimen@i
    \true@sw
   }{%
    \false@sw
   }%
  }%
 }%
 \outputdebug@sw{\saythe\dimen@\saythe\dimen@i\saythe\dimen@ii}{}%
\@ifdim{\ht\z@=\z@}{%
\@ifdim{\ht\tw@=\z@}{%
\true@sw
}{%
\false@sw
}%
}{%
\true@sw
}%
{%
}{%
\ltxgrid@info{Unsatifactorily balanced columns: giving up}%
\setbox\tw@\box#1%
\setbox\z@ \box#2%
}%
 \setbox\tw@\vbox{\unvbox\tw@\vskip\z@skip}%
 \setbox\z@ \vbox{\unvbox\z@ \vskip\z@skip}%
 \set@colroom
\dimen@\ht\z@\@ifdim{\dimen@<\ht\tw@}{\dimen@\ht\tw@}{}%
\@ifdim{\dimen@>\@colroom}{\dimen@\@colroom}{}%
 \outputdebug@sw{\saythe{\ht\z@}\saythe{\ht\tw@}\saythe\@colroom\saythe\dimen@}{}%
\setbox#1\vbox to\dimen@{\unvbox\tw@\unskip\raggedcolumn@skip}%
\setbox#2\vbox to\dimen@{\unvbox\z@ \unskip\raggedcolumn@skip}%
\outputdebug@sw{{\tracingall\scrollmode\showbox#1\showbox#2}}{}%
}%
\catcode`\@=12\relax

\usepackage{graphicx}
\usepackage{dcolumn}
\usepackage{amsmath}
\usepackage{epsfig}
\usepackage{subfigure}

\input pubboard/babarsym

\newcommand{\BABARPubYear}    {08}
\newcommand{\BABARPubNumber}  {032}
\newcommand{\SLACPubNumber} {13352}
\newcommand{\LANLNumber} {xxxxx}
\def\kStar{{\bar K}^{*0}}
\def\dPhiGamma {D^{0} \rightarrow \phi \gamma}
\def\dRhoGamma {D^{0} \rightarrow \rho^{0} \gamma}
\def\dOmegaGamma {D^{0} \rightarrow \omega \gamma}
\def\dKsGamma{D^{0} \rightarrow K^0_{S} \gamma}
\def\dKStarGamma{D^{0} \rightarrow \kStar \gamma}
\def\dKStarPi{D^{0} \rightarrow \kStar \pi^{0}}
\def\dKstarPhiGamma{D^{0} \rightarrow V \gamma,\,V=\kStar, \phi}
\def\dKPiPi{D^{0} \to K^{-} \pi^{+} \pi^{0}}
\def\dKStarEta{D^{0} \to \kStar \eta}
\def\lumi{387.1 \invfb}
\def\dKPi {D^{0} \rightarrow K^{-} \pi^{+}}
\def\dStar{D^{*+}}

\def\dStarTag{\dStar \to D^{0} \piSlow}

\def\phiKK{\phi \rightarrow K^{-} K^{+}}

\def\rhoPiPi{\rho^{0} \to \pi^{-} \pi^{+}}

\def\piGG{\pi^{0} \rightarrow \gamma \gamma}

\def\kStar{{\bar K}^{*0}} 
\def\kShort{K_{S}^{0}}

\def\dKPiPi{D^{0} \rightarrow K^{-} \pi^{+} \pi^{0}}
\def\dPhiPi{D^{0} \rightarrow \phi \pi^{0}}
\def\dKsPi{D^{0} \rightarrow \kShort \pi^{0}}

\def\dPhiEta {D^{0} \rightarrow \phi \eta}
\def\dKsEta{D^{0} \rightarrow \kShort \eta}

\def\dVGamma{D^{0} \rightarrow V \gamma}
\def\dVPi{D^{0} \rightarrow V \pi^{0}}

\def\dStarTag{D^{*+} \rightarrow D^{0} \pi^{+}}

\def\nPhiGamma{242.6 \pm 24.8}
\def\nKStarGamma{2285.8 \pm 113.2}
\def\nKPi{(335.1 \pm 4.0) \times 10^{3}}

\def\effPhiGamma{(10.8 \pm 0.1)\%}
\def\effKStarGamma{(6.4 \pm 0.1) \%}
\def\effKPi{(5.3 \pm 0.2) \%}

\newcommand{\dKStarGammaRelVal}{8.43}
\newcommand{\dKStarGammaRelStatErr}{0.51}
\newcommand{\dKStarGammaRelSysErr}{0.70}

\newcommand{\dKStarGammaAbsVal}{3.22}
\newcommand{\dKStarGammaAbsStatErr}{0.20}
\newcommand{\dKStarGammaAbsSysErr}{0.27}

\newcommand{\dPhiGammaRelVal}{7.15}
\newcommand{\dPhiGammaRelStatErr}{0.78}
\newcommand{\dPhiGammaRelSysErr}{0.69}

\newcommand{\dPhiGammaAbsVal}{2.73}
\newcommand{\dPhiGammaAbsStatErr}{0.30}
\newcommand{\dPhiGammaAbsSysErr}{0.26}

\newcommand{\belleResultTot}{(2.43^{+0.66}_{-0.57}(stat.)^{+0.12}_{-0.14}(sys.)) \times 10^{-5}}

\newcommand{\dKStarGammaRelResult}{(\dKStarGammaRelVal \pm \dKStarGammaRelStatErr \pm \dKStarGammaRelSysErr) \times 10^{-3}}
\newcommand{\dKStarGammaAbsResult}{(\dKStarGammaAbsVal \pm \dKStarGammaAbsStatErr \pm \dKStarGammaAbsSysErr) \times 10^{-4}}
\newcommand{\dPhiGammaRelResult}{(\dPhiGammaRelVal \pm \dPhiGammaRelStatErr \pm \dPhiGammaRelSysErr) \times 10^{-4}}
\newcommand{\dPhiGammaAbsResult}{(\dPhiGammaAbsVal \pm \dPhiGammaAbsStatErr \pm \dPhiGammaAbsSysErr) \times 10^{-5}}

\def\figurebox#1#2#3{%
    \def\arg{#3}%
    \ifx\arg\empty
    {\hfill\vbox{\hsize#2\hrule\hbox to #2{\vrule\hfill\vbox to #1{\hsize#2\vfill}\vrule}\hrule}\hfill}%
    \else
    {\hfill\epsfbox{#3}\hfill}%
    \fi}

\long\def\inst#1{\par\nobreak\kern 4pt\nobreak
    {\it #1}\par\vskip 10pt plus 3pt minus 3pt}

\begin{document}

\begin{flushleft}
hep-ex/\LANLNumber\\
SLAC-PUB-\SLACPubNumber\\
\babar-PUB-\BABARPubYear/\BABARPubNumber\\
\end{flushleft}

\title{
{\large \bf
Measurement of the Branching Fractions of the Radiative Charm Decays $\dKStarGamma$ and $\dPhiGamma$ } 
}

%
\author{B.~Aubert}
\author{M.~Bona}
\author{Y.~Karyotakis}
\author{J.~P.~Lees}
\author{V.~Poireau}
\author{E.~Prencipe}
\author{X.~Prudent}
\author{V.~Tisserand}
\affiliation{Laboratoire de Physique des Particules, IN2P3/CNRS et Universit\'e de Savoie, F-74941 Annecy-Le-Vieux, France }
\author{J.~Garra~Tico}
\author{E.~Grauges}
\affiliation{Universitat de Barcelona, Facultat de Fisica, Departament ECM, E-08028 Barcelona, Spain }
% ...
\collaboration{The \babar\ Collaboration}
\noaffiliation

\date{\today}

\begin{abstract}
We present a measurement of the branching fractions
for the Cabibbo-favored radiative decay, $\dKStarGamma$,
and the Cabibbo-suppressed radiative decay, $\dPhiGamma$.
These measurements are based on a data
sample corresponding to an integrated luminosity of $\lumi$,
recorded with the \babar\ detector at the PEP-II
$e^+e^-$ asymmetric-energy collider operating at center-of-mass
energies 10.58 and 10.54 GeV.
We measure the branching fractions
relative to the well-studied decay $\dKPi$ and find
${\cal B}(\dKStarGamma)/{\cal B}(\dKPi) = \dKStarGammaRelResult$ and
${\cal B}(\dPhiGamma)/{\cal B}(\dKPi) = \dPhiGammaRelResult$,
where the first error is statistical and the second is systematic. 
This is the first measurement of  ${\cal B}(\dKStarGamma)$.
\end{abstract}


\maketitle
\begin{figure}[h]
  \begin{center}
    \includegraphics[width = \columnwidth]{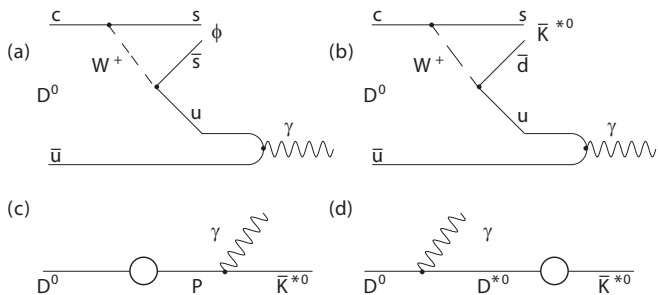}
    \caption[Long-range contributions]{Feynman diagrams for the long-range
      electromagnetic contributions to $\dKstarPhiGamma$. 
      Figures (a) and (b) show sample vector dominance processes,
      while (c) and (d) are examples of pole diagrams, where the circles
    signify the weak transition and P represents a pseudoscalar meson.  
     }
    \label{fig:d0KStarGamma_LD}
  \end{center}
\end{figure}

In the $b$-quark sector, radiative decay processes have provided a
rich field in which to
study the Standard Model of particle physics. 
Decays such as $B \rightarrow \rho \gamma$
have yielded measurements of the Cabibbo-Kobayashi-Maskawa
matrix element $|V_{td}|$ \cite{Ali:2001ez, Ali:2004hn}. 
These decays are dominated by short-range
electroweak processes, whereas long-range contributions are
suppressed.
The situation is reversed in the charm sector, 
where radiative decays are expected to be
dominated largely by non-perturbative processes, 
examples of which are shown schematically in Fig. 
\ref{fig:d0KStarGamma_LD}.  
Long-range
contributions to radiative charm decays 
are expected to increase the branching fractions 
for these modes to values of the order of $10^{-5}$, 
whereas short-range interactions are 
predicted to yield rates at the $10^{-8}$ level.
Given the expected dominance of long-range 
processes, radiative charm decays provide 
a laboratory in which to test these QCD-based calculations.\\
\indent
Numerous theoretical models have
been developed to describe these radiative charm decays 
\cite{Bajc:1994ui, Bajc:1995ys, golowich, Cheng:1994kp, fajfer, PhysRevD564302, Fajfer:1998dv}.
The two most comprehensive studies
\cite{golowich, Fajfer:1998dv} predict very similar
amplitudes for the dominant diagrams shown 
in Fig. \ref{fig:d0KStarGamma_LD}.  The first paper bases 
predictions on Vector Meson Dominance (VMD) calculations, 
while the second paper uses Heavy-Quark Effective Theory
in conjunction with Chiral-Lagrangians. Though 
each approach arrives at similar estimates for the 
magnitudes of the individual decay amplitudes, Ref. \cite{golowich} 
predicts that the pole diagrams, shown in Figs. \ref{fig:d0KStarGamma_LD} (c) and (d),
interfere destructively and cancel nearly completely.  Ref. \cite{Fajfer:1998dv} makes no such 
predictions.
Precise measurements of ${\cal B}(\dKstarPhiGamma)$ may provide insight into the amount of interference 
between pole diagrams.  
\\
\indent
The first observation of a radiative, but color-suppressed, 
$D^{0}$ decay process was made by the Belle collaboration with a measurement of
${\cal B}(\dPhiGamma) = \belleResultTot$ \cite{belle}.
CLEO II conducted searches for other radiative decays and established the current upper limit of
${\cal B}(\dKStarGamma) <~ 7.6 \times 10^{-4}$ at $90\%$ confidence level (C.L.),
as well as upper limits on ${\cal B}(\dRhoGamma)$ and ${\cal B}(\dOmegaGamma)$ \cite{cleo}.
Table \ref{tbl:searchSummary} summarizes theoretical predictions and current experimental results.\\
\begin{table}[floatfix]
  \begin{center}
    \begin{tabular}{lcc}
      \hline
      Mode & Experimental &  Theoretical\cite{Bajc:1994ui, Bajc:1995ys, golowich, Cheng:1994kp, fajfer, PhysRevD564302, Fajfer:1998dv} \\
      & B.F. $(\times 10^{-5})$ & B.F. $(\times 10^{-5})$\\
      \hline
      $\dPhiGamma$ & $(2.43^{+0.66}_{-0.57}(stat.)^{+0.12}_{-0.14}(sys.)$ \cite{belle} & $0.1 - 3.4$ \\
      $\dKStarGamma$ & $< 76$ (90\% C.L.) \cite{cleo} & $7 - 80$ \\
      $\dRhoGamma$ & $< 24$ (90\% C.L.) \cite{cleo} & $0.1 - 6.3$ \\
      $\dOmegaGamma$ & $< 24$ (90\% C.L.) \cite{cleo} & $0.1 - 0.9$ \\
      \hline
    \end{tabular}
    \caption{The current experimental status and theoretical
    predictions for the branching fractions (B.F.) of radiative charm
    decays with vector mesons.}
    \label{tbl:searchSummary}
  \end{center}
\end{table}
\indent
In this paper we present the first observation of the Cabibbo-favored radiative decay $\dKStarGamma$, as 
well as an improved branching fraction measurement of the previously observed decay $\dPhiGamma$.
The analysis is based on $\lumi$ of data recorded by the \babar\
detector at the PEP-II $e^+e^-$ asymmetric-energy collider operating
at center-of-mass (CM) energies of $\sqrt{s} = 10.58 \gev$ and $10.54 \gev$,
and uses approximately $5 \times 10^{8}~ e^{+}e^{-} \rightarrow c\overline{c}$ events.

\begin{figure}[htbp]
  \centering
  \subfigure[\label{fig:phiGamma_D0MassFit_Data} The $\phi\gamma$ invariant mass distribution.]
            {\includegraphics[width = 0.69\columnwidth]{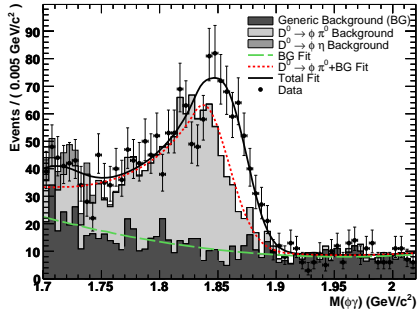}}\\
  \subfigure[\label{fig:phiGamma_HelicityFit_Data} The $\phi\gamma$ helicity angle distribution.]
            {\includegraphics[width = 0.69\columnwidth]{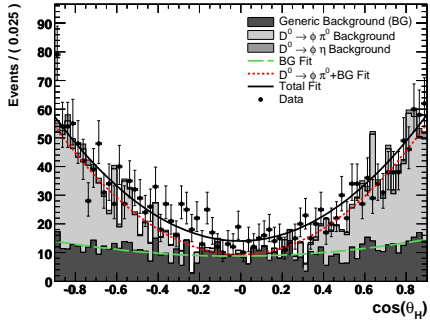}}\\
  \subfigure[\label{fig:totD0MassFit_Data} The $\kStar\gamma$ invariant mass distribution.]
            {\includegraphics[width = 0.69\columnwidth]{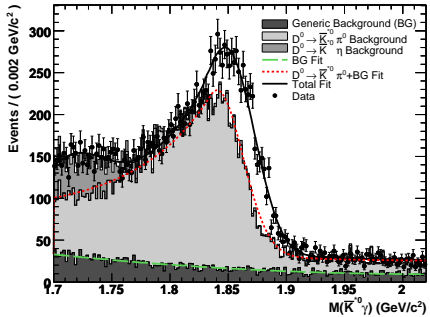}}\\
  \subfigure[\label{fig:nKStarGamma_Helicity} The $\dKStarGamma$ helicity angle distribution.]
            {\includegraphics[width = 0.69\columnwidth]{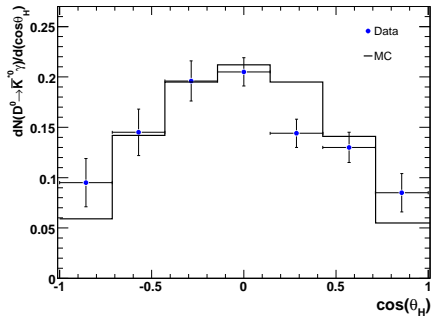}}
  \caption{Invariant mass and $\cos \theta_{H}$ distributions for data (points) and simulated events (histograms).
    The curves show the fit results and the individual signal and background contributions.
 BG refers to the combinatoric background.}
\label{fig:dataFits}%
\end{figure}

The \babar\ detector is described in detail 
elsewhere~\cite{detector}.  Charged particle momenta 
are measured with a 5-layer double-sided silicon 
vertex tracker and a 40-layer drift chamber.  
Charged hadron identification is provided by measurements 
of the specific ionization energy loss, $dE/dx$, in the 
tracking system and of the Cherenkov angle obtained from 
a ring-imaging Cherenkov detector.  An 
electromagnetic calorimeter consisting of 6580 
CsI(Tl) crystals measures shower energy and position for 
electrons and photons. 
These detector elements are located inside, and coaxial with, the 
cryostat of a superconducting solenoidal magnet, which provides
a 1.5 T magnetic field.
The instrumented flux return of the magnet allows 
discrimination between muons and pions.
\\
\indent
A detailed Monte Carlo (MC) simulation of the \babar\ detector 
based on \geant4~\cite{geant4} is used
to validate the analysis and determine 
the reconstruction efficiencies.
We optimize our selection criteria using 
simulated events by maximizing significance,
defined as $N_{S}/\sqrt{N_{S}+N_{B}}$, where 
$N_S$ and $N_B$ denote
the number of signal and background candidates
in the MC simulation.
\indent
We reconstruct radiative
$\dKstarPhiGamma$
decays using the charged decay modes of the
vector meson, $\kStar \to K^{-} \pi^{+}$ 
($\phi \to K^{-} K^{+}$) \cite{charge}.
We form $\kStar$ ($\phi$) candidates from pairs of 
oppositely charged tracks identified as $K^-\pi^+$ ($K^{-}K^{+}$) 
using the Cherenkov angle measurement 
of the DIRC and $dE/dx$
measurements from the tracking system,
and accept any $K^{-} \pi^{+}$ ($K^{-} K^{+}$) candidates with 
invariant mass in the range 
$0.848$ to $0.951 \gevcc$ ($1.01$ to $1.03 \gevcc$).
The charged track candidates are fit
to a common vertex, and 
a fit probability greater than $0.1 \%$ is required.
\\
\indent
A photon candidate is defined as an energy deposit in the EMC
that is not associated with the trajectory of a charged track,
and which exhibits the expected shower shape characteristics.
Each such candidate is required to have CM energy greater than
0.54 GeV. The charged-particle vertex is assumed to be the
production point of the photon.
We suppress the significant background from
$\piGG$ decays by rejecting a photon candidate which, when paired with
another photon in the event, 
results in an invariant mass consistent with the  $\pi^{0}$ mass, 
$(0.115 < M(\gamma \gamma) < 0.150) \gevcc$. 
\\
\indent
Background from random $\dVGamma$ candidates
is reduced by
requiring that the $D^{0}$ candidate be a product of the decay
$\dStarTag$.  A $\dStar$ candidate is formed by combining
a $D^{0}$ candidate with a low-momentum charged pion,
denoted as $\pi_{s}^{+}$.  These pion candidates are required to have  
CM momentum less than $450 \mevc$.  We calculate the mass difference,
$\Delta M = M(V \gamma \pi_{s}^{+}) - M(V \gamma)$ and require
$(0.1435 < \Delta M < 0.1475)\gevcc$. 
The $\Delta M$ distribution of candidates arising from signal 
decays is well-described by a Gaussian distribution function.  
Our selection corresponds to a six-standard deviation interval centered on the mean of
the Gaussian, and hence retains almost all of the signal
candidates. 
We reduce combinatoric background from $B\bar{B}$ events to a negligible level by requiring that the CM momentum of
the $D^{*+}$ candidate be greater than $2.62 \gevc$.\\
\indent
The dominant background in our sample of $\dKStarGamma$ 
candidates results from  $\dKPiPi$ decays, where
one of the photons from the $\pi^0$ decay is paired with the kaon and
pion from the $D^{0}$ decay to closely mimic the signal mode.
As described above, we use a $\pi^0$ veto to suppress such events but, given
the large branching fraction of this mode,
${\cal B}(\dKPiPi) = (13.5 \pm 0.6)\%$ \cite{pdg}, 
a significant number of such candidates survives. We can
separate this background from signal on a statistical basis 
because of differences in the
$K^-\pi^+\gamma$ invariant mass distribution. The background
distribution peaks slightly below the nominal $D^0$ mass, and has a
different shape from that of signal events.
An additional background arises from $\dKStarEta$ events where the $\eta$
decays to two photons, one of which is combined with the $K^{-} \pi^{+}$ pair
to form an invariant mass within our $D^{0}$ mass window. This contribution 
peaks well below the nominal $D^0$ mass,
and it can be separated easily from correctly reconstructed $\dKStarGamma$ decays.
\\
\indent
The impact of both $\dKStarPi$ and $\dKStarEta$ is further reduced by using
the $\kStar$
helicity angle $\theta_H$.  The helicity
angle is defined as the angle between the momentum of the $\kStar$ meson parent particle
($D^{0}$) and the  momentum of the $\kStar$ daughter kaon as measured in the $\kStar$ rest
frame.  Due to angular momentum conservation, $dN/d\cos \theta_{H}$ 
for signal candidates varies as
$1-\cos ^{2}\theta_{H}$, whereas for $D^{0} \rightarrow \kStar \pi^0(\eta)$
events the cosine of the helicity angle is $\cos^{2}\theta_{H}$ distributed.
The $\cos\theta_{H}$ distribution of $\dKPiPi$ candidates is complicated by the interference and overlap
of resonant structure in the final state Dalitz plot.
Based on a MC study
an asymmetric selection of $-0.30<\cos \theta_H<0.65$ is
chosen to maximize the signal significance.
\\
\indent
Similarly, but to a lesser extent, the signal of the Cabibbo-suppressed radiative decay $\dPhiGamma$ 
is obscured by backgrounds from $\dPhiPi$ and $\dPhiEta$ decays.
Due to the small width of the $\phi$ meson, background from 
$D^{0} \to K^{-} K^{+} \pi^{0}$ transitions with a $K^+K^-$ invariant mass in the $\phi$ region yields a negligible 
contribution to $\dPhiGamma$ \cite{bill}.  
Since angular momentum conservation dictates that the cosine of the helicity angle of the remaining $\dPhiPi$ events follow 
 a $\cos ^{2}\theta_{H}$ distribution, 
we replace the tight $\cos \theta_{H}$ selection criterion used in the $\dKStarGamma$ case
with the looser requirement $|\cos \theta_{H}| < 0.9$ and include $\cos \theta_H$ as a variable in the 
fitting procedure.
This retains a larger fraction of 
signal events, and so reduces statistical uncertainty.
\\
\indent
We consider other radiative decays which might reflect into the 
$M(\kStar \gamma)$ and $M(\phi \gamma)$ invariant mass distributions.  Background to 
$\dKStarGamma$ may arise from $\dPhiGamma$ if a kaon from 
$\phiKK$ is mis-identified as a pion.  Background from $\dRhoGamma$
may arise if a pion from $\rhoPiPi$ is misidentified as a kaon.  
Real $\dKStarGamma$ events can reflect into the 
$M(\phi \gamma)$ distributions if a $\pi^{+}$ is misidentified 
as a $K^{+}$.  Using MC simulations, all of these background contributions are
found to be negligible.\\
\indent
We extract the $\dKStarGamma$ yield 
using an unbinned extended maximum likelihood
method (E-MLM) to fit the $M(\kStar \gamma)$ invariant mass 
spectrum.  The yield of $\dPhiGamma$ events is 
extracted using an E-MLM to fit the two dimensional 
distribution of invariant mass, $M(\phi \gamma)$, and helicity, $\cos \theta_{H}$.  
\\
\indent
We use a Crystal Ball (CB) line shape \cite{cry} to model the invariant mass distributions for
$\dKStarGamma$ ($\dPhiGamma$) signal events, and background reflections 
from $\dKPiPi$ ($\dPhiPi$) decays.
The invariant mass distributions 
of $\dKStarEta$ and $\dPhiEta$ background
events are modeled with a Gaussian function and a first
order Chebychev polynomial. The remaining combinatoric background decays 
are modeled with a second order Chebychev polynomial.
In the $\phi\gamma$, case the $\cos \theta_{H}$ distributions of $\dPhiGamma$, $\dPhiPi$, 
$\dPhiEta$, and combinatoric background events are all modeled using second 
order Chebychev polynomials.
The parameters of these probability distribution functions (PDFs) are obtained using
simulated events and subsequently fixed when fitting the data.
\\
\indent
We validate the invariant mass PDFs using data.
To verify that the MC correctly simulates the backgrounds and
the effects of the missing photon from the $\pi^0$ decay, we search our
data sample for  $D^0\rightarrow K^0_S\gamma$  candidates. 
Since this decay is forbidden by angular momentum conservation, 
the candidates surviving our selection criteria are 
all combinatoric background or due to $D^0 \rightarrow K^0_S\pi^0$ or $D^0
\rightarrow K^0_S\eta$ decays.
\\
\indent
We select $K^0_S$ candidates from pairs of oppositely charged tracks identified as pions.  
The pions are required to share a common production vertex and have an
invariant mass in the range $(0.490 < M(\pi^{+} \pi^{-}) < 0.505) \gevcc$.
Selection criteria for the photon momentum, $\dStar$ momentum, $\Delta M$, and $\pi^{0}$ 
veto are identical to those used
in the $\dVGamma$ analyses. The resulting ${K^0_S \gamma}$ invariant mass spectrum is fit with
a linear combination of three PDFs.
The first PDF is used to model $\dKsPi$ candidates, and has the same
functional form as the one used to model $\dKPiPi$ candidates.
The second PDF is used to model $\dKsEta$ candidates, and has the same
functional form as that used to model $\dKStarEta$ candidates.
The third PDF is a second order Chebychev polynomial used to model combinatoric background candidates.  The
shapes for both $\dKsEta$ and combinatoric background candidates are fixed using MC. The $\dKsPi$ signal shape
is allowed to float in the final fit.  Both MC and data are fit in
this way and we find good agreement. The observed differences in the fit parameters are used to correct 
the CB line shape PDFs as described below.
\\
\indent
A second test is performed using $\dKStarGamma$ candidates taken from the sideband regions defined by
$|\cos \theta_{H}| > 0.9$.  Very few
$\dKStarGamma$ candidates are seen within this region, leading to a clean sample of $\dKPiPi$ decays.
The resulting $D^{0}$ invariant mass spectrum is fit using a procedure
similar to the one used for signal region $\dKStarGamma$ candidates.
The only differences are that the $\dKStarGamma$ contribution is
fixed to zero, and the $\dKPiPi$ signal
shape is allowed to float freely. The $M(\kStar \gamma)$ distribution 
of $\dKPiPi$ events is compared between data and MC and we find good agreement.
\\
\indent
Potential differences between the $\dVGamma$ invariant mass
distributions for data and MC are evaluated by using $\dKsPi$ events.
The selection criteria for $K^0_S$ mesons are identical to those
applied in the $\dKsGamma$ analysis. The requirements on $\Delta M$
and $\dStar$ CM momentum are as before.
A $\pi^0$ candidate consists of a photon pair
with invariant mass satisfying $(0.110 < M(\gamma\gamma) < 0.150) \gevcc$,
and resultant laboratory momentum greater than $0.540 \gevc$.
This resulting sample of $D^0\rightarrow K^0_S\pi^0$ candidates is fit
to a CB line shape and a linear background.
\\
\indent
We used the average difference between the CB line shape parameters in MC and these data control samples to modify the 
PDF parameterizations used in the fit.
\\
\indent
The fit results from data and expected signal and background contributions from MC are shown in Fig. \ref{fig:dataFits}(a-c). 
The event yields obtained from the E-MLM fit for both $\dPhiGamma$ and 
$\dKStarGamma$ are $N(\dPhiGamma;~ \phi \to K^{-} K^{+}) = \nPhiGamma$ and 
$N(\dKStarGamma;~ \kStar \to K^{-} \pi^{+}) = \nKStarGamma$.
The reconstruction efficiencies,
determined using MC, are found to be 
$\epsilon(\dPhiGamma;~ \phi \to K^{-} K^{+}) = \effPhiGamma$ and 
$\epsilon(\dKStarGamma;~ \kStar \to K^{-} \pi^{+}) = \effKStarGamma$.
\\
\indent
In order to avoid uncertainties in the overall normalization we measure the branching fraction of the radiative decays
relative to ${\cal B}(\dKPi)$.  We prepare a $\dKPi$ dataset following procedures similar to those described 
above, and find a yield
$N(\dKPi) = \nKPi$ with an efficiency of $\epsilon(\dKPi) = \effKPi$.
\\
\indent
We perform several consistency checks.
Our result is compared to the $\cos \theta_{H}$ distribution expected for $\dKStarGamma$ by refitting the data
in intervals of $\cos \theta_{H}$ and measuring $N(\dKStarGamma)$ in each interval. 
The normalized and efficiency-corrected result, shown in Fig. \ref{fig:nKStarGamma_Helicity},
compares well to the expected distribution.
As an additional check, we divide the data into five distinct samples, one for each PEP-II run period, and
perform the analysis on each subset independently.
We see a $\dKStarGamma$ signal for each run period, and find that the branching ratios are
consistent within statistical uncertainties.\\
\indent
We evaluate the systematic uncertainties associated with our
measurement in several different studies.
Systematic effects due to the PDF parameterizations of signal and
backgrounds are determined by generating an ensemble
of 1,000 random numbers drawn from a normal distribution for each PDF
parameter, including their correlations obtained from our fits.
We refit the data using each of the $1,000$
sets of random numbers. The resulting distribution of $N(\dVGamma)$ 
is fit to a Gaussian function and the percent standard deviation is 
taken as the systematic error, $ 5.9\% $ for $\dPhiGamma$ and 
$4.4 \%$ for $\dKStarGamma$.
\\
\indent
Correcting the $\dVGamma$ and $\dVPi$ PDF parameters 
using the data control samples induces a second systematic uncertainty in the parameterization 
of the signal shapes.  We estimate this effect by independently applying the corrections 
obtained using each of the three control samples.
The largest percentage variation  in $N(\dVGamma)$ is taken as the systematic uncertainty associated with this correction;
this leads to systematic uncertainities of $3.0 \%$ for $N(\dPhiGamma)$ and $4.3 \%$ for 
$N(\dKStarGamma)$.
\\
\begin{linenomath}
\begin{table}[htbp]
  \begin{center}
    \begin{tabular}{lcc}
      \hline
      Systematic & $ \sigma(\dPhiGamma)~(\%)$ & $\sigma(\dKStarGamma)~(\%)$ \\
      \hline
      Tracking, vertexing & $1.2 $ & $1.0$ \\
      Particle ID & $2.9$ & $1.1$ \\
      $\gamma$ reconstruction & $1.8$ & $1.8$ \\
      $\pi^{0}$ veto & $1.8$ & $1.8$ \\
      PDF parameter & $5.9$ & $4.4$ \\
      Correcting ${\cal P}_{\dVGamma}$  & $3.0$ & $4.3$ \\
      and ${\cal P}_{\dVPi}$ & &  \\
      Ref. mode efficiency & $1.5$ & $1.5$ \\
      Selection criteria & $5.4$ & $4.5$ \\
      \hline
      Total systematic effect & $9.6$ & $8.3$ \\
      \hline
    \end{tabular}
    \caption[Summary of all systematic errors]
            {Summary of all systematic errors for each $D^0$ decay mode. 
	      The total systematic uncertainty is obtained by adding the individual systematic
              estimates in quadrature.}\label{tbl:sysSummary}
  \end{center}
\end{table}
\end{linenomath}
\indent
We quantify the difference in particle identification (PID) efficiency between data and simulation by means of a
high-purity control sample of $D^{*+}\to D^0 \pi^+$, $\dKPi$ events, which we
divide into intervals of polar angle and momentum. The change in yield when
PID selection criteria are applied is computed separately for data and
for simulated events and the difference is taken as a correction factor
for that interval. We then weight the correction factors according to the expected
momentum and polar-angle distributions of the $\dKStarGamma$ signal.  
While a portion of the PID systematic uncertainty for our signal modes 
is canceled when measuring the branching fractions in ratio to $\dKPi$, 
the residual uncertainty is found to be $2.88 \%$ for $\dPhiGamma$ and 
$1.10 \%$ for $\dKStarGamma$.  By measuring ${\cal B}(\dKStarGamma)$ 
and ${\cal B}(\dPhiGamma)$ with respect to $\dKPi$, first-order effects from charged particle tracking
also cancel, leaving only a second order systematic uncertainty of $1.00 \%$ for $\dKStarGamma$ 
events and $1.20 \%$ for $\dPhiGamma$.  We summarize all systematic uncertainties
in Table \ref{tbl:sysSummary}.\\
\indent
In this paper, we report our observation of the
Cabibbo-favored, but color-suppressed, radiative decay $\dKStarGamma$.  
We also present confirmation of the previous 
measurement of the Cabibbo-suppressed radiative decay
${\cal B}(\dPhiGamma)$, but with reduced statistical uncertainities.
The measured branching ratios are 
\begin{linenomath}
\begin{eqnarray*}
\frac{{\cal B}(\dPhiGamma)}{{\cal B}(\dKPi)} & = & \dPhiGammaRelResult \label{eq:brRel_PhiGamma}\\
\frac{{\cal B}(\dKStarGamma)}{{\cal B}(\dKPi)} & = & \dKStarGammaRelResult \label{eq:brRel_KStarGamma}
\end{eqnarray*}
\end{linenomath}
where the first uncertainty is statistical and the second is systematic.
\indent
Using the current 
world average of ${\cal B}(\dKPi) = (3.82 \pm 0.07) \%$ \cite{pdg} we obtain the following absolute branching fractions:
\begin{linenomath}
\begin{eqnarray*}
{\cal B}(\dPhiGamma) & = & \dPhiGammaAbsResult \label{eq:brAbs_PhiGamma}\\
{\cal B}(\dKStarGamma) & = & \dKStarGammaAbsResult. \label{eq:brAbs_KStarGamma}
\end{eqnarray*}
\end{linenomath}
These results are consistent with the theoretical expectations of Table \ref{tbl:searchSummary}.
\\
\indent
In the context of the vector dominance model the largest contribution to radiative $D^0$ decays is expected to come from
a virtual $\rho^0$ coupling directly to a single photon, leading to the prediction that the branching ratios
${\cal B}(\dPhiGamma)/{\cal B}(\dKStarGamma)$ and
${\cal B}(D^{0} \to \phi \rho^{0})/{\cal B}(D^{0} \to \kStar \rho^{0})$ should be equal \cite{golowich}. 
Comparing our measurements of the radiative $D^0$ decays with the current world averages \cite{pdg} we find
\begin{linenomath}
\begin{eqnarray*}
  \frac{{\cal B}(\dPhiGamma)}{{\cal B}(\dKStarGamma)}
  & = & (6.27 \pm 0.71 \pm 0.79) \times 10^{-2} \label{eq:compareVGamma}\\
  && \nonumber \\
  \frac{{\cal B}(D^{0} \to \phi \rho^{0})}{{\cal B}(D^{0} \to \kStar \rho^{0})}
  & = & (6.7 \pm 1.6) \times 10^{-2} \label{eq:compareVRho}
\end{eqnarray*}
\end{linenomath}
in agreement with this prediction.\\
\indent
If we assume all contributions are from VMD type processes and
under the assumption that the $\rho^{0}$ meson is transversely polarized, as has 
been confirmed experimentally for $D^{0} \to \kStar \rho^{0}$ \cite{pdg}, we expect 
${\cal B}(\dVGamma) \approx \alpha_{EM} {\cal B}(D^{0} \to V \rho^{0})$ \cite{golowich},
where $\alpha_{EM} = 1/137$ is the fine structure constant.
Using our results we find
\begin{linenomath}
\begin{eqnarray*}
{\cal B}(\dKStarGamma) & = &(0.021 \pm 0.005)\;{\cal B}(D^{0} \to \kStar \rho^{0})\\
{\cal B}(\dPhiGamma)  &=& (0.020 \pm 0.003)\;{\cal B}(D^{0} \to \phi \rho^{0})
\end{eqnarray*}
\end{linenomath}
which in both cases is about a factor of three larger than the VMD prediction.
This indicates that we are seeing enhancements from
processes other than VMD, which might be explained by incomplete cancellation
between pole diagrams.\\
\indent
We are grateful for the excellent luminosity and machine conditions
provided by our \pep2\ colleagues, 
and for the substantial dedicated effort from
the computing organizations that support \babar.
The collaborating institutions wish to thank 
SLAC for its support and kind hospitality. 
This work is supported by
DOE
and NSF (USA),
NSERC (Canada),
CEA and
CNRS-IN2P3
(France),
BMBF and DFG
(Germany),
INFN (Italy),
FOM (The Netherlands),
NFR (Norway),
MES (Russia),
MEC (Spain), and
STFC (United Kingdom). 
Individuals have received support from the
Marie Curie EIF (European Union) and
the A.~P.~Sloan Foundation.

\bibliography{ref}

\begin{thebibliography}{17}
\expandafter\ifx\csname natexlab\endcsname\relax\def\natexlab#1{#1}\fi
\expandafter\ifx\csname bibnamefont\endcsname\relax
  \def\bibnamefont#1{#1}\fi
\expandafter\ifx\csname bibfnamefont\endcsname\relax
  \def\bibfnamefont#1{#1}\fi
\expandafter\ifx\csname citenamefont\endcsname\relax
  \def\citenamefont#1{#1}\fi
\expandafter\ifx\csname url\endcsname\relax
  \def\url#1{\texttt{#1}}\fi
\expandafter\ifx\csname urlprefix\endcsname\relax\def\urlprefix{URL }\fi
\providecommand{\bibinfo}[2]{#2}
\providecommand{\eprint}[2][]{\url{#2}}

\bibitem[{\citenamefont{Ali and Parkhomenko}(2002)}]{Ali:2001ez}
\bibinfo{author}{\bibfnamefont{A.}~\bibnamefont{Ali}} \bibnamefont{and}
  \bibinfo{author}{\bibfnamefont{A.~Y.} \bibnamefont{Parkhomenko}},
  \bibinfo{journal}{Eur. Phys. J.} \textbf{\bibinfo{volume}{C23}},
  \bibinfo{pages}{89} (\bibinfo{year}{2002}).

\bibitem[{\citenamefont{Ali et~al.}(2004)\citenamefont{Ali, Lunghi, and
  Parkhomenko}}]{Ali:2004hn}
\bibinfo{author}{\bibfnamefont{A.}~\bibnamefont{Ali}},
  \bibinfo{author}{\bibfnamefont{E.}~\bibnamefont{Lunghi}}, \bibnamefont{and}
  \bibinfo{author}{\bibfnamefont{A.~Y.} \bibnamefont{Parkhomenko}},
  \bibinfo{journal}{Phys. Lett.} \textbf{\bibinfo{volume}{B595}},
  \bibinfo{pages}{323} (\bibinfo{year}{2004}).

\bibitem[{\citenamefont{Bajc et~al.}(1995)\citenamefont{Bajc, Fajfer, and
  Oakes}}]{Bajc:1994ui}
\bibinfo{author}{\bibfnamefont{B.}~\bibnamefont{Bajc}},
  \bibinfo{author}{\bibfnamefont{S.}~\bibnamefont{Fajfer}}, \bibnamefont{and}
  \bibinfo{author}{\bibfnamefont{R.~J.} \bibnamefont{Oakes}},
  \bibinfo{journal}{Phys. Rev.} \textbf{\bibinfo{volume}{D51}},
  \bibinfo{pages}{2230} (\bibinfo{year}{1995}).

\bibitem[{\citenamefont{Bajc et~al.}(1996)\citenamefont{Bajc, Fajfer, and
  Oakes}}]{Bajc:1995ys}
\bibinfo{author}{\bibfnamefont{B.}~\bibnamefont{Bajc}},
  \bibinfo{author}{\bibfnamefont{S.}~\bibnamefont{Fajfer}}, \bibnamefont{and}
  \bibinfo{author}{\bibfnamefont{R.~J.} \bibnamefont{Oakes}},
  \bibinfo{journal}{Phys. Rev.} \textbf{\bibinfo{volume}{D54}},
  \bibinfo{pages}{5883} (\bibinfo{year}{1996}).

\bibitem[{\citenamefont{Burdman et~al.}(1995)\citenamefont{Burdman, Golowich,
  Hewett, and Pakvasa}}]{golowich}
\bibinfo{author}{\bibfnamefont{G.}~\bibnamefont{Burdman}},
  \bibinfo{author}{\bibfnamefont{E.}~\bibnamefont{Golowich}},
  \bibinfo{author}{\bibfnamefont{J.~L.} \bibnamefont{Hewett}},
  \bibnamefont{and} \bibinfo{author}{\bibfnamefont{S.}~\bibnamefont{Pakvasa}},
  \bibinfo{journal}{Phys. Rev.} \textbf{\bibinfo{volume}{D52}},
  \bibinfo{pages}{6383} (\bibinfo{year}{1995}).

\bibitem[{\citenamefont{Cheng et~al.}(1995)}]{Cheng:1994kp}
\bibinfo{author}{\bibfnamefont{H.-Y.} \bibnamefont{Cheng}}
  \bibnamefont{et~al.}, \bibinfo{journal}{Phys. Rev.}
  \textbf{\bibinfo{volume}{D51}}, \bibinfo{pages}{1199} (\bibinfo{year}{1995}).

\bibitem[{\citenamefont{Fajfer et~al.}(2003)\citenamefont{Fajfer, Prapotnik,
  Prelovsek, Singer, and Zupan}}]{fajfer}
\bibinfo{author}{\bibfnamefont{S.}~\bibnamefont{Fajfer}},
  \bibinfo{author}{\bibfnamefont{A.}~\bibnamefont{Prapotnik}},
  \bibinfo{author}{\bibfnamefont{S.}~\bibnamefont{Prelovsek}},
  \bibinfo{author}{\bibfnamefont{P.}~\bibnamefont{Singer}}, \bibnamefont{and}
  \bibinfo{author}{\bibfnamefont{J.}~\bibnamefont{Zupan}},
  \bibinfo{journal}{Nucl. Phys. Proc. Suppl.} \textbf{\bibinfo{volume}{115}},
  \bibinfo{pages}{93} (\bibinfo{year}{2003}).

\bibitem[{\citenamefont{Fajfer and Singer}(1997)}]{PhysRevD564302}
\bibinfo{author}{\bibfnamefont{S.}~\bibnamefont{Fajfer}} \bibnamefont{and}
  \bibinfo{author}{\bibfnamefont{P.}~\bibnamefont{Singer}},
  \bibinfo{journal}{Phys. Rev.} \textbf{\bibinfo{volume}{D56}},
  \bibinfo{pages}{4302} (\bibinfo{year}{1997}).

\bibitem[{\citenamefont{Fajfer et~al.}(1999)\citenamefont{Fajfer, Prelovsek,
  and Singer}}]{Fajfer:1998dv}
\bibinfo{author}{\bibfnamefont{S.}~\bibnamefont{Fajfer}},
  \bibinfo{author}{\bibfnamefont{S.}~\bibnamefont{Prelovsek}},
  \bibnamefont{and} \bibinfo{author}{\bibfnamefont{P.}~\bibnamefont{Singer}},
  \bibinfo{journal}{Eur. Phys. J.} \textbf{\bibinfo{volume}{C6}},
  \bibinfo{pages}{471} (\bibinfo{year}{1999}).

\bibitem[{\citenamefont{Abe~{\it et al.}}(2004)}]{belle}
\bibinfo{author}{\bibfnamefont{K.}~\bibnamefont{Abe~{\it et al.}}},
  \bibinfo{journal}{Phys. Rev. Lett.} \textbf{\bibinfo{volume}{92}},
  \bibinfo{pages}{101803} (\bibinfo{year}{2004}), \bibinfo{note}{the published
  result has been rescaled using the latest values from [15].}

\bibitem[{\citenamefont{Asner~{\it et al.}}(1998)}]{cleo}
\bibinfo{author}{\bibfnamefont{D.~M.} \bibnamefont{Asner~{\it et al.}}},
  \bibinfo{journal}{Phys. Rev.} \textbf{\bibinfo{volume}{D58}},
  \bibinfo{pages}{092001} (\bibinfo{year}{1998}).

\bibitem[{\citenamefont{Aubert~{\it et al.}}(2002)}]{detector}
\bibinfo{author}{\bibfnamefont{B.}~\bibnamefont{Aubert~{\it et al.}}},
  \bibinfo{journal}{Nucl. Instrum. Meth.} \textbf{\bibinfo{volume}{A479}},
  \bibinfo{pages}{1} (\bibinfo{year}{2002}).

\bibitem[{\citenamefont{Agostinelli et~al.}(2003)}]{geant4}
\bibinfo{author}{\bibfnamefont{S.}~\bibnamefont{Agostinelli}}
  \bibnamefont{et~al.}, \bibinfo{journal}{Nucl. Instrum. Meth.}
  \textbf{\bibinfo{volume}{A506}}, \bibinfo{pages}{250} (\bibinfo{year}{2003}).

\bibitem[{cha()}]{charge}
\bibinfo{note}{Unless explicitly stated otherwise, charge conjugate reactions
  are included throughout this paper.}

\bibitem[{\citenamefont{Yao~{\it et al.}}(2006)}]{pdg}
\bibinfo{author}{\bibfnamefont{W.-M.} \bibnamefont{Yao~{\it et al.}}}
  (\bibinfo{collaboration}{Particle Data Group}), \bibinfo{journal}{J. Phys.}
  \textbf{\bibinfo{volume}{G33}}, \bibinfo{pages}{1} (\bibinfo{year}{2006}),
  \bibinfo{note}{and 2007 partial update for the 2008 edition.}

\bibitem[{\citenamefont{Aubert~{\it et al.}}(2007)}]{bill}
\bibinfo{author}{\bibfnamefont{B.}~\bibnamefont{Aubert~{\it et al.}}},
  \bibinfo{journal}{Phys. Rev.} \textbf{\bibinfo{volume}{D76}},
  \bibinfo{pages}{011102} (\bibinfo{year}{2007}).

\bibitem[{\citenamefont{{Oreglia, Ph.D Thesis, SLAC-236}}(1980)}]{cry}
\bibinfo{author}{\bibfnamefont{M.~J.} \bibnamefont{{Oreglia, Ph.D Thesis,
  SLAC-236}}} (\bibinfo{year}{1980}), \bibinfo{note}{{J. E.} Gaiser, Ph.D.
  Thesis, SLAC-255 (1982), T. Skwarnicki, Ph.D Thesis, DESY F31-86-02 (1986)}.

\end{thebibliography}

\end{document}